\documentclass[12pt,epsfig]{article}

\usepackage{amsmath,amssymb}

\usepackage{latexsym,graphicx,epsfig}

\usepackage{wrapfig}

\date{}

\newcommand{\bm}{\boldmath}

\newcommand{\be}{\begin{equation}}

\newcommand{\ee}{\end{equation}}

\newcommand{\bear}{\be\begin{array}}

\newcommand{\bea}{\begin{eqnarray}}

\newcommand{\eea}{\end{eqnarray}}

\newcommand{\dst}{\displaystyle}

\newcommand{\fr}[2]{\dfrac{{\dst #1}}{{\dst #2}}}

\def\lsi{\raise0.3ex\hbox{$<$\kern-0.75em\raise-1.1ex\hbox{$\sim$}}}

\def\gsi{\raise0.3ex\hbox{$>$\kern-0.75em\raise-1.1ex\hbox{$\sim$}}}

\newcommand{\lsim}{\mathop{\lsi}}

\def\Pom{{\bf I\!P}}

\def\bu{$\bullet$\,}

\def\bes{\begin{subequations}}

\def\ees{\end{subequations}}

\newcommand{\fn}[1]{\footnote{{ #1}}}

\def\nn{\nonumber}

\title {\bf \bm How to measure the Pomeron phase
in diffractive dipion photoproduction}

\author{{\em I.F. Ginzburg$^{1}$\thanks{email: ginzburg@math.nsc.ru},
I.P. Ivanov$^{1,2}$\thanks{email: igivanov@cs.infn.it}}\\
{\small $^1$ Sobolev Institute of Mathematics, Novosibirsk, Russia}\\
{\small $^2$ INFN, Gruppo Collegato di Cosenza, Cosenza, Italy}}

\begin{document}

\maketitle

\abstract{The study of charge asymmetry of pions in the
high-energy process $e p\to e\pi^+\pi^-p$ ($\gamma p\to\pi^+\pi^-
p$) at very small dipion momenta offers a method to measure the
phase of the forward hadronic (quasi)elastic amplitude $\gamma
p\to\rho p$. We estimate potential of such measurements at HERA.}

\section{Introduction}\label{Intro}

The phase $\delta_F$ of the forward amplitude of the hadronic
elastic scattering
 \be
{\cal A}=|{\cal A}|e^{i\delta_F}\equiv |{\cal A}| \exp\left[i
{\pi\over 2}(1+\Delta_F)\right] \label{def0}
 \ee
at high energy, treated often as a Pomeron phase, 
is an important object in hadron physics.

However, the object, studied in modern experiments and dubbed the
Pomeron, seems to be complex. In some models it is the same for
all processes, in other models it is process-dependent, which manifests
itself in different effective intercepts in different processes.
The measurement of the phase of this object in various
processes will be a useful step towards clarification of its
nature. For example, in the naive Regge-pole Pomeron model, this
phase is related directly to the Pomeron intercept,
$\Delta_F=-(\alpha_\Pom-1)$, in the model of a dipole Pomeron,
$\Delta_F=-(\alpha_\Pom -1) - \pi/(2\ln (s/s_0))$, \cite{jenk},
for the model with Regge pole and cuts one adds to the value given by
Pomeron pole intercept the contribution of the branch cut with
process dependent coefficient.

Up to the moment, a phase of such type was measured  at very high
energy only for $pp$, $\bar{p}p$ elastic scattering (via the study
of Coulomb interference near forward direction, see latest results
\cite{pp} and references to earlier experiments therein). Such
experiments demand detailed measurement of the cross section at
extremely low transverse momentum of recorded particle,
$p_\bot\approx\sqrt{|t|} \lsim 30$ MeV, which translates into very
small scattering angles.

\bu Here we propose to measure a similar phase for the process
$\gamma p\to\rho p$ via the study of charge asymmetry of pions in
the diffractive process
 \be
ep\to e\pi^+\pi^- p\qquad (\gamma p\to\pi^+\pi^-
p)\,.\label{process}
 \ee
To describe our proposal in more detail, we denote by $p_\pm$ the
momenta of the $\pi^\pm$ and
 \be
r^\mu = p^\mu_+ - p^\mu_-\,,\quad k^\mu = p^\mu_+ + p^\mu_-\,,
\quad M=\sqrt{k^2}\,.\label{pion}
  \ee
We propose to measure charge asymmetry of pions in reaction
(\ref{process}) in the region
 \be
(20\div 30)\mbox{ MeV}< k_\bot< 100\;\mbox{ MeV}\,,\quad
1.1<M<1.4\mbox{ GeV }\,.
 \label{dom0}
  \ee
Essential part of our description is valid also at
$M<1.1$~GeV. However, we do not present here definite
predictions for experiment for this mass region due to
complex structure of the $C$-even amplitude of dipion
production here. At $M>1.4$ GeV diffractive exclusive
production of pion pairs becomes a too rare process to use
it in the considered problems.

The main mechanism of the reaction in this domain is
diffractive photoproduction of dipions in the $C$-odd
state (the $\rho$--meson and its "tails", including
$\rho'$) via the "physical Pomeron" --- the vacuum quantum
number exchange in $t$--channel. The phase of the {\it
amplitude of this "physical Pomeron"} (\ref{def0}) is the main
subject of our study. Besides, dipions can be produced in
the $C$--even state via ({\it i}) the $\rho,\,\omega$ Regge
exchange, ({\it ii}) the odderon exchange and ({\it iii})
the one-photon exchange with proton (the Primakoff effect).

The interference of amplitudes of the $C$-odd and $C$-even dipion
production provides charge asymmetry of the observed pion
distribution. The experimental study of this charge asymmetry is a
good tool for the investigation of a number of phenomena
\cite{chargereview}.

These exchanges have very different dependence on the
transverse momentum of dipion $k_\bot$.

The Primakoff effect is strongly peaked at small transverse
momenta of dipion $k_\bot$. It can be neglected at
$k_\bot>200$ MeV (see details in the text below). It is
natural to expect that the odderon contribution, just as the
$\rho/\omega$ Reggeon exchanges, has a flatter $k_\bot$
dependence, similarly to other hadronic amplitudes. Besides, the
contribution of the $\rho/\omega$ Reggeon exchanges was
estimated as very small in the HERA energy region
\cite{GIN2001}, and according to modern data the odderon
contribution is low enough, so that both these contributions
can be neglected at the considered low transverse momenta
(\ref{dom0}). Therefore, (see \cite{GIv2004})

{\it (i)}\; At $k_\bot<100$ MeV the charge asymmetry discussed is
described by an interference of the Pomeron contribution with the
Primakoff one, and it is sensitive to the phase $\delta_F$. This
sensitivity offers a method to measure $\delta_F$ in the discussed
experiments.

{\it (ii)}\; At $k_\bot\approx 0.3-1$ GeV the Primakoff
contribution is negligible, the discussed charge asymmetry is
described by an interference of the Pomeron and odderon
contributions, and this very experiment provides an opportunity to
discover the odderon \cite{GIN2001}, \cite{GIv2004}.

$\bullet$ {\it Proposed experimental set-up}. We suggest to
observe dipion final state without other particles in
detectors (without observation of scattered proton or
electron). The pions that hit the detector have transverse
momenta $p_{\pm\bot}\sim M/2 \sim 500$ MeV with emission
angles $20\div 150$ mrad, which looks not so difficult to
measure. It is the sum of the transverse momenta of the two
pions $k_\bot$ that is supposed to be  small and
measurable. So, in order for this method to be efficient,
we need a reasonable resolution of the reconstruction of
each pion's transverse momentum. The choice of the lower
bound in $k_\bot$ (\ref{dom0}) corresponds to the
anticipated accuracy of this measurement.

Let us stress a vital feature of our suggestion. The
procedure we propose {\em does not} demand the measurement
of very small scattering angles of pions.

The quality of this set-up can be controlled via measurement of
charge symmetric part of cross section (CSP) by two ways. First,
the observation of events with the same pion content and recording
of scattered electron with $k_{\bot e}\le 30\div 50$ MeV has low
efficiency. However these observations will give  CSP in the
considered kinematical region with good enough accuracy. Second,
the known results for the CSP at higher total transverse momentum
(obtained with recording of electron and proton) can be used for
extrapolation in the kinematical region under interest.

\bu In the next section we discuss kinematics of the process and
introduce the charge asymmetric variables. In Section 3 we present
well known amplitudes of $C$--odd and $C$--even dipion production.
In Section 4 we study the differential cross section and find the
integral charge asymmetries. In Section 5 we present numerical
results for $\gamma p$ and $ep$ collisions. Discussion and conclusions 
are found in Section 6.

\section{Kinematics }

In the proposed set-up without recording of electrons, the
main contribution to the $ep\to e\pi^+\pi^- p$ cross
section is given by convolution of equivalent photon
spectrum with cross section of mass shell $\gamma
p\to\pi^+\pi^- p$ subprocess. In the considered region of
$k_\bot$ accuracy of this equivalent photon (or
Weiszacker--Williams) approximation is very high (much
better than $k_{\bot max}^2/m_\rho^2$). We stress again that 
the procedure we propose
{\em does not} demand the measurement of very small
scattering angles of pions. Therefore we focus first on
the subprocess -- the dipion quasielastic photoproduction
off proton $\gamma p\to \pi^+\pi^- p$ considering the
limitation in $k_\bot$ \eqref{dom0} as that for this
subprocess. The convolution with equivalent photon spectrum
is considered in sec.~\ref{secep}.

{\bf\bm Process  $\mathbf \gamma p\to\pi^+\pi^- p$}. The
energies we have in mind correspond to the HERA energy
range ($\sqrt{s_{\gamma p}}\; \sim\; 100 \div 200$ GeV).
The initial momenta of the photon and proton are $q$ and
$P$ respectively, $s=(q+P)^2$, initial photon polarization
vector is $\vec{e}$. We use kinematical variables
(\ref{pion}) for this process as well.

We define the $z$-axis as the $\gamma p$ collision axis and label
the vectors orthogonal to this axis by $\bot$. Let us denote by
$z_+$ and $z_-$ the standard light cone variables for each charged
pion, $z_\pm\approx (\epsilon_\pm+p_{\pm z})/(2E_\gamma)=(p_\pm
P)/(qP)$ (for the considered process $z_++z_-=1$).

We direct the $x$-axis along vector $\vec k_\bot$ and define by
$\psi$ the azimuthal angle of the linear photon polarization with
respect to the fixed lab frame of reference. For instance, for the
photon in electroproduction $e p \to e \pi^{+}\pi^{-}p$ virtual
photons are polarized in the electron scattering plane and
$\psi$ is the azimuthal angle relative to the electron
scattering plane. Then the polarization vector of the initial
photon with helicity $\lambda_\gamma=\pm 1$ can be written as
${\bf\bm\vec e}^{\lambda} = -{1 \over\sqrt{2}}\cdot
e^{-i\lambda_\gamma \psi} (\lambda_\gamma, i)$.

It is useful also to consider polar and azimuthal angles of
$\pi^+$ in the dipion c.m.s, $\theta$ and $\phi$, and the
velocity of a pion in this frame $\beta =
\sqrt{1-4m_\pi^2/M^2}$, so that $r_{c.m.s.}= $ $\beta
M(0,\,\sin\theta\cos\phi\,,\;\sin\theta\cos\phi\,,\;
cos\theta)$. We denote by $J$ the total angular momentum
(total spin) of dipion, by $\lambda_\gamma$ and
$\lambda_{\pi^+\pi^-}$ the helicities of photon and
produced dipion, respectively, and by $n=
|\lambda_\gamma-\lambda_\rho|$ -- the value of helicity
flip for each amplitude. (The final results are averaged
over initial photon polarizations.) Instead of phase
analysis in terms of these angular variables, many physical
problems can be solved definitely via the measurement of
charge asymmetry of pions.

The phenomenon of charge asymmetry is the difference in
distributions of particles and antiparticles. It is determined by
the part of differential cross section that changes its sign under
$r^\mu\to -r^\mu$ change. Particularly, we describe the {\bf
forward--backward (FB) and transverse (T) asymmetries} by
variables
 \bear{lcl}
FB:&\quad\xi=\fr{z_+-z_-}{\beta (z_++z_-)}
&=\cos\theta\,,\\[3mm]
T:& \quad v= \fr{{p}_{+\bot}^2-{p}_{-\bot}^2-\xi {
k}_\bot^2}{\beta M|{ k}_\bot|}\equiv\fr{({\rho}_\bot {
k}_\bot)}{\beta M|{ k}_\bot|}&=\sin\theta\cos\phi\;; \quad
{\rho}_\bot={r}_\bot-\xi {k}_\bot\,.\end{array}
 \label{asyms}\ee

We consider the amplitude of the dipion production ${\cal A}$,
which is normalized so that
 \be
d\sigma = |{\cal A}|^2 \ \beta dM^{2}\, d k^{2}_\bot\, d\cos\,
\theta\, d\phi\,{d\psi \over 2\pi} = {2 \over \sqrt{1-\xi^2-v^2}}\
|{\cal A}|^2\ \beta dM^{2}\, d k^{2}_\bot\,d\xi\,dv\,{d\psi \over
2\pi}\;.\label{phsp}
 \ee
We will see below that only transverse asymmetry arises in the
considered case. We describe the values of this charge asymmetry
and the charge symmetric background by quantities
 \be
\Delta\sigma_T = \int d\sigma(v >0) - \int d\sigma(v <0)\,, \quad
\sigma_{bkgd}=\int d\sigma\,, \label{Delsig}
 \ee
with integration over (identical) suitable region of final phase
space.

\section{The amplitudes}

Note first that in the considered range of momentum transfer
(\ref{dom0}) the inelastic transitions in the proton vertex as
well as helicity-flip elastic transitions are small.

$\bullet$ {\bf \bm The $C$-odd dipion diffractive production} is
described by the "physical Pomeron". It has been studied both in
theory and in experiment (e.g. at HERA) as a production of $C$--odd
resonances, mainly $\rho(770)$ meson with well known properties.

Our basic assumption is that the amplitude can be written in a
form
 \be
 {\cal A}= \sum\limits_{Jn}A_{Jn}(s,t,M^2)D_J(M^2){\cal
 E}_J^{\lambda_\gamma,\lambda_{\pi\pi}}\;.\label{genampl}
 \ee

{\it (i)} The first factor $A_{Jn}(s,t,M^2)$ is "the Pomeron
amplitude" for the production of the dipion state with effective
mass $M$, angular momentum $J$ and helicity flip $n$. In the
considered mass region the contribution with $J=1$ dominates (the
admixture of $J=3$ looks negligible). In discussions we assume
that in the considered mass interval the entire dependence of
amplitude ${\cal A}$ on dipion mass $M$ at $t\approx 0$ can be
accumulated with high precision in the factor $D_J(M^2)$ so that the
amplitude $A_{Jn}$ is only weakly dependent on the dipion mass $M$.
It is normalized in the $\rho$--meson peak in such a manner that
$A_{1n=0}(M=M_\rho)= e^{i\delta_F}|A_{1n}|$ with
 \be
|A_{1n}|^2=\sigma_\rho Be^{-B k_\bot^2}\qquad\mbox{ with }
B\approx 10\mbox{ GeV}^{-2}\,,\;\;\sigma_\rho\approx 11\,\mu b
\,.\label{rhocrsec}
 \ee

The $s$-channel helicity conservation $\lambda_{\pi^+\pi^-} =
\lambda_\gamma$ for process (\ref{process}) is a well established
fact. For the considered $k_\bot$ region, the helicity violating
amplitudes are as small as \ $\sim(|k_\bot|/M)^n\le (0.03)^n$, and
we neglect them below. Last, in the considered region (\ref{dom0})
the $t$--dependence of the amplitude is negligible.

{\it (ii)} The second factor $D_J(M)$ describes the decay of this
dipion state to pions --- it is driven by strong interaction of
pions in the final state. In similarity to construction of the pion 
formfactor, It should be constructed from
contributions of $\rho$, $\rho^{\prime}$, $\rho^{\prime\prime}$
in a manner to describe
data in the effective mass interval considered . At
$2m_\pi<M<M_\rho$ one can use for $D_1$ the well known
Gounaris--Sakurai approximation obtained for the pion
form--factor. At $M>M_\rho$ one should take into account also
$\rho'$, etc. states with variable parameters given by coupling
constants and parameters of $\rho'$, $\rho''$. The reasonable
parameterization should give complete description of dipion mass
spectrum $\propto |D_1|^2$ in the considered region\fn{ To take
into account possible dependence $B(M)$, one should distinguish
here a quantity defined at $t=0$ and that obtained by
extrapolation procedure, the second quantity can contain also
factor appearing due to $t$--integration of $e^{-B_Mk_\bot^2}$.
These quantities will be obtained from two different methods of
verification of the proposed set-up (see end of sec.~\ref{Intro}).
The first method gives the quantity at $t\approx 0$, the second
method CAN include integration over $t$.}. Below, we use the fit
from \cite{DM2}. It covers the required mass interval and includes
$\rho$ running width and $\rho'/\rho''$ states. Note that the
parameters of the model can be fixed with a better accuracy with
detailed measurement of the dominant $C$--even dipion mass spectrum
in the very experiment we propose.

The qualitative discussion becomes transparent with the standard
Breit--Wigner factor for $R=\rho(770)$ (including $R\pi^+\pi^-$
coupling)
 \be
D_J(M^2) \approx  {\sqrt{m_R\Gamma_R Br(R\to\pi^+\pi^-)/\pi} \over
-M^2 +m_R^2 - im_R\Gamma_R}\,.\label{BW}
 \ee

{\it (iii)}  The third factor ${\cal
E}^{\lambda_\gamma,\lambda_{\pi\pi}}_J$ describes the angular
distribution of pions in their rest frame, ${\cal
E}^{\lambda_\gamma,\lambda_{\pi\pi}}_J =
Y^{J,\lambda_{\pi\pi}}(\theta,\phi)e^{-i\lambda_\gamma\psi}$.

Finally, the amplitude of the $C$--odd dipion photoproduction reads
as
 \bear{c}
{\cal A}_-= e^{i\delta_F}  \cdot |A_{1,0}(s)| \cdot
D_{1}(M^2)\cdot \sqrt{\dfrac{3}{ 8\pi}} \sin\theta
e^{i\lambda_\gamma(\phi-\psi)}\\[3mm]
\equiv e^{i\delta_F}  \cdot |A_{1,0}(s)| \cdot D_{1}(M^2)\cdot
\sqrt{\dfrac{3}{ 8\pi}} \sqrt{1-\xi^2}\,
e^{i\lambda_\gamma(\phi-\psi)} \,.\end{array} \label{basampdef1}
 \ee

Here and below subscript $-$ or $+$ at ${\cal A}$ denotes the
$C$-parity of the produced dipion.

$\bullet$ {\bf \bm The amplitude of the production of $C$-even
dipions via Primakoff effect} is the same as that in the
two--photon processes $e^+e^-\to e^+e^-\pi^+\pi^-$
\cite{1975,GSS2001}. In the regions under interest (\ref{dom0})
the dominant contribution is given by the almost real photon
exchanges with both electron and proton. Therefore, the total
helicity of the initial two--photon state and respectively of
dipions can be 0 and 2. The amplitude can be written in a form
similar to eq.~(\ref{genampl}).

Beginning from the threshold, the pions interact strongly in the
$I=J=0$ state (which is described by $f_0$ resonances). The other
partial waves are described well with the QED approximation for
point--like pions (with known small modifications). The QED
amplitude with $I=0$, $J=2$ become surprisingly large starting
from $M=0.5-0.7$ GeV. The other amplitudes can be neglected
everywhere in our problem.

At $M^2\ll s_{\gamma p}$ the amplitude  of the Primakoff $\gamma
p\to R p$\ \  process can be written \cite{1975} via the
two-photon decay width $\Gamma^R_{\gamma\gamma}$ of the resonance
$R$ with spin $J$
 \bes\label{agam} \be
 {A}_\gamma= \sqrt{\sigma_2} \cdot {|{k}_\bot| \over {k}_\bot^2
+ Q^2_m}\;\mbox{ with }\;Q_m^2=\left(\dfrac{m_p M^2}{
s}\right)^2\,,\quad \sigma_2 \equiv {8
\pi\alpha\Gamma^R_{\gamma\gamma} (2J+1)\over
m_R^3}\,.\label{agamma}
 \ee
Here, $Q^2_m$ is the minimal value of the virtuality of the
exchanged photon (typically $Q_m^2<m_e^2$\ ).

At $1.1<M<1.4$ GeV, the main contribution is given by the $I=0$,
$J=2$ partial wave, other partial waves being negligible. Here
$f_2(1270)$--meson ($J=2$) production dominates. We define by
$g_0$ and $g_2$ the relative probability amplitudes of the dipion
production in the states with helicity 0 and 2; $g_0^2+g_2^2=1$.
According to data, the contribution of total helicity
$\lambda_{\pi\pi}= 2$ dominates, i.e. $g_2\gg |g_0|$ (see e.g.
\cite{gamgampipi}). Similarly to (\ref{basampdef1}), the amplitude
of the process can be written as
 \bear{c}
{\cal A}_+= A_\gamma \cdot D_{2}(M^2)\cdot \left( g_2
Y_{2,2}(\theta,\phi) + g_0 Y_{2,0}(\theta,\phi)
\right)e^{-i\lambda_\gamma\psi}\,,\\[3mm]
\equiv A_\gamma \cdot D_{2}(M^2)\cdot
\sqrt{\dfrac{15}{32\pi}}\left[g_2(1-\xi^2)e^{2i
\lambda_\gamma\phi} + g_0\sqrt{{2\over 3}}(3\xi^2
-1)\right]e^{-i\lambda_\gamma\psi}\,.
\end{array}\label{basampdef2}
 \ee\ees
with $D_2$ factor given by eq.~\eqref{BW} with $R=f_2(1270)$.

For more precise calculation the QED contribution should be
also accounted for. In the unitarized model describing the
data near the $f_2$ peak constructed in \cite{f2shape} the
$I=0,\, J=2$ partial wave contains a phase-shifted
Breit-Wigner factor and a modified Born QED term: 
\be
D_2(M^2) = {\sqrt{m_f\Gamma_f(M^2) Br(f_2\to\pi^+\pi^-)/\pi}
\over -M^2 +m_f^2 - im_f\Gamma_f(M^2)} \cdot e^{i\zeta} +
D^{QED}_2(M^2)\,. \label{D2corr} 
\ee 
The mass dependence of
the $f_2$ width $\Gamma_f(M^2)$ and the parametrization for
the modified QED contribution $D^{QED}_2(M^2)$ were taken
from \cite{f2shape}. The phase factor $e^{i\zeta}$ represents
one particular possibility to effectively fulfil the
unitarity constraint: the value of $\zeta$ is such that $D_2$
becomes purely imaginary at $M = m_f$.

\section{Cross sections}

\subsection{Differential cross section. Photoproduction.}

The differential cross section of the ${\gamma
p\to\pi^+\pi^-p}$ sub-process at\linebreak[4]
$1.1<M<1.4$~GeV averaged over the initial photon
polarizations is
 \bea
&d \sigma = 2\fr{ |{\cal A}_- + {\cal A}_+|^2}{
\sqrt{1-\xi^2-v^2}} \beta dM^2 d{k}_\perp^2 d\xi dv
 =d\sigma_{sym}+d\sigma_{asym}\,,&\label{sigtot}\\
 &\fr{d\sigma_{sym}}{ dM^2 d{k}_\perp^2 d\xi dv} = \fr{2 \beta}{
\sqrt{1-\xi^2-v^2}}\left\{ |A_{1,0}(s)|^2 |D_1(M^2)|^2 \fr{3}{
8\pi} (1 -\xi^2) + \right.&\label{secPombckg}\\
 &\left.\fr{15}{16 \pi} A_\gamma^2 |D_2|^2
\left[\fr{g_2^2}{2}(1 -\xi^2)^2 +3g_0^2
(\xi^2-{1\over 3})^2
 +g_0g_2\sqrt {6}(\xi^2-{1\over 3})
(2v^2+\xi^2-1)\right]\right\}\,,& \label{secphotbckg}\\[3mm]
&\begin{array}{c} \fr{d\sigma_{asym}}{dM^2 d{k}_\perp^2 d\xi dv} =
v \cdot \fr{\beta}{\sqrt{1-\xi^2-v^2}} \cdot \fr{3\sqrt{5}}{ 4
\pi}
 |A_{1,0}(s)|\, A_\gamma \cdot Re\left[D_1 e^{i\delta_F}
D_2^\dagger\right]\\[3mm]
  \cdot \left[g_2(1-\xi^2) + g_0\sqrt{\fr{2}{3}}
(3\xi^2-1)\right]\,.\end{array} \label{dif1}&
 \eea

Here $d\sigma_{sym}$ represents the charge-symmetric contribution,
which comes from the squares of the Pomeron and of the Primakoff
amplitudes. The interference between these amplitudes produces the
charge asymmetric contribution $d\sigma_{asym}$. Since the phase
$\delta_F$ enters only $d\sigma_{asym}$, we need to extract charge
asymmetry, for this task the charge symmetric contribution
$d\sigma_{sym}$ is a background.

The appearance of factor $v$, {\it describing transverse charge
asymmetry}, in the interference term is very natural. First, due
to integration over $\psi$, we are left with terms diagonal in
photon polarization states, i.e. $\lambda_\gamma$ is the same in
${\cal A}_\pm$ and in ${\cal A}^\dagger_\pm$. Then, ${\cal
A}_-{\cal A}_+^*$ can be rewritten as a charge symmetric factor
multiplied by $\sin\theta e^{\pm i\lambda_\gamma\phi}$. The
averaging over initial photon polarizations means that we sum
contributions with opposite helicities, i.e. consider the sum
which is proportional to $\sin\theta e^{\pm i\lambda_\gamma\phi}
+h.c.\Rightarrow v $. In other words, the averaging over photon
polarization transforms complex factors from the spherical
harmonics to the real factor describing charge asymmetry.

For our case when one can consider only one partial wave in the
Primakoff amplitude, the $M$ dependence in $d\sigma_{asym}$ is
described completely by the {\it overlap function}, which is
independent on  the $g_0$ and $g_2$ interrelation,
 \be
{\cal I}_{\rho f}(M^2) = Re\left[ D_1 e^{i\delta_F}
D_2^\dagger\right]\,.\label{overlap}
 \ee

\begin{figure}[htb]
   \centering
\includegraphics[width=0.8\textwidth]
{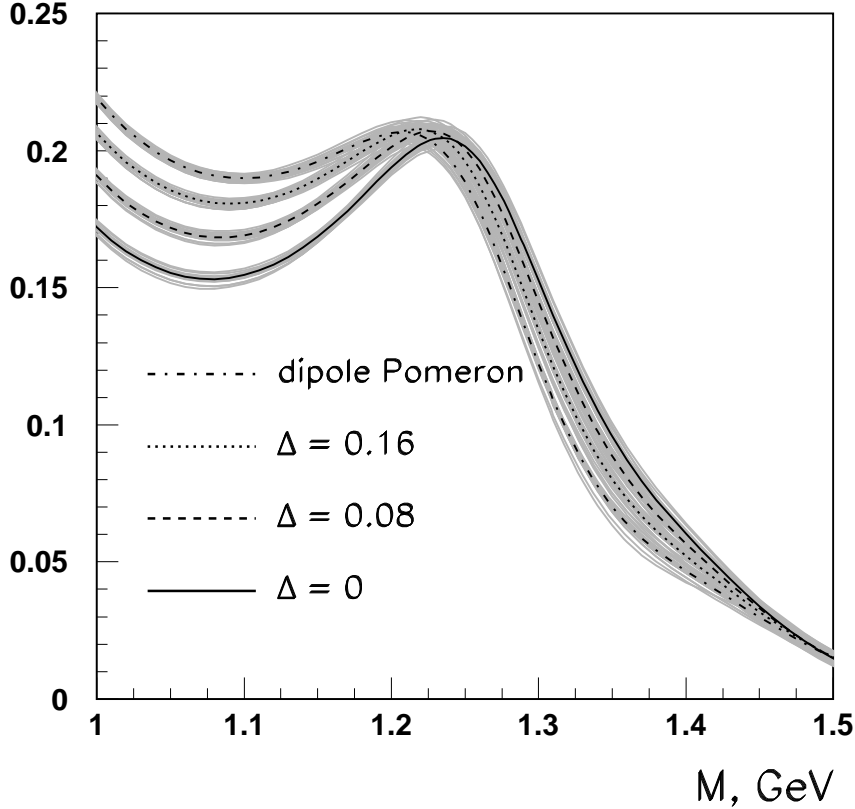}
   \caption{\sf Overlap function ${\cal I}_{\rho f}$ vs. $M^2$.
The black solid, dashed, and dotted curves correspond to
simple pole Pomeron model with $\Delta_F = $ 0, 0.08, and
0.16, respectively. The dashed-dotted curve represents one
particular parametrization of a dipole Pomeron model,
\cite{jenk}, calculated for $\sqrt{s_{\gamma p}} = 50$ GeV.
In each of the four cases, the grey region 
corresponds to 1$\sigma$ variations in the
parameters of $\rho'$, $\rho''$ and $f_2$ resonances.}
   \label{fig1}
\end{figure}

{\large\bf The shape of $M$-dependence}. Fig.~\ref{fig1}
demonstrates the overlap functions for several Pomeron models.
The parameterization for $D_1$ was taken from \cite{DM2}
(with $\rho/\rho'/\rho''$ parameters and running of the $\rho$ width
taken into account), while the parametrization
for $D_2$, which included both the $f_2$ resonance and the
$I=0,\, J=2$ modified born term, was taken from
\cite{f2shape}. The four black curves correspond here to
different Pomeron models. The solid, dashed, and dotted
curves correspond to the simple pole Pomeron model with
$\Delta_F = 0$, 0.08, and 0.16, respectively. 
The dashed-dotted curve represents one
particular parametrization of a dipole Pomeron model,
\cite{jenk}, calculated for $\sqrt{s_{\gamma p}} = 50$ GeV.
In each of the four cases, the grey region
corresponds to 1$\sigma$ variations of the parameters in
$D_i$ used. The resulting shaded region allows one to see
the typical level of inaccuracy that arises from the
parameterizations used. It is not large, and allows one to
discern different Pomeron models for $M_{\pi\pi}$
interval below the $f_2$ peak.

Note that many qualitative  features can be easily understood in
the simplified $\rho$--meson model for $D_1$ (\ref{BW}). It was
found numerically that this approximation gives also reasonable
quantitative approximation for the overlap function.

{\large\bf Dependence on the momentum transfer}. Integrating the
differential cross section (\ref{Delsig}) over the whole $\xi, v$
space, within mentioned $M$ interval, and with $k_\bot$ interval
$k_{max}>k_\bot>k_{min}$ (\ref{dom0}), one obtains:
 \be
\sigma_{bkgd}= \sigma_\rho B C_1({k}_{max}^2 - {k}_{min}^2)
+\sigma_2C_2\ln {{k}_{max}^2 \over
{k}_{min}^2+Q^2_m}\,,\;\;C_i=\int dM^2
|D_i(M^2)|^2\,,\label{bkgd1}
 \ee

Numerical estimates show that the second term here, which
represents the Primakoff contribution, can be neglected at
considered values $k_{max}$ and $k_{min}$ (\ref{dom0}). The
integral value of the charge asymmetry, $\Delta\sigma_{chas,T}$,
calculated in the same $M$ and $k_\bot$ regions, is
 \be
\Delta\sigma_{chas,T} = {9\sqrt{5} \over 8} \sqrt{\sigma_\rho B
\cdot \sigma_2} \cdot \Delta{\cal I}_{\rho f}\cdot({k}_{max} -
{k}_{min}) \,,\quad\Delta{\cal I}_{\rho f}=\int dM^2{\cal I}_{\rho
f}(M^2)\,.\label{asym1}
 \ee
To obtain simple estimates, we set $g_2 = 1,\, g_0=0$.

\subsection{\boldmath $ep$ collisions}\label{secep}

We think that the most efficient
way to study the problem under interest  is to investigate dipion
production in $ep\to e\pi^+\pi^-p$, e.g. at HERA without recording
of scattered electron and proton (and without other particles in
detector except $\pi^+$ and $\pi^-$).

The $ep$ cross section is given by a convolution of the virtual
photon flux originating from the electron with the cross section
of the $\gamma p$ subprocess. The dominant part of the $ep$ cross
section comes from region of very small virtuality of the emitted
photon. That is the base of the equivalent photon approximation
(see e.g. \cite{1975}), in which the flux of the equivalent
photons with energy $\omega=yE_e$ and transverse momentum $q_\bot$
is
 \be
dn_\gamma = {\alpha \over
\pi}\dfrac{dy}{y}\left[1-y+\dfrac{y^2}{2}-(1-y)\dfrac{q_e^2}{q_\bot^2}\right]
 { q_\bot^2 d q_\bot^2 \over
(q_\bot^2 + q_e^2)^2} \;\;\mbox{ with }\; q_e^2={m_e^2y^2\over
1-y}\,.\label{WWflux}
 \ee
Note that the photon energy $\omega$ coincides with the total
dipion energy with high accuracy.

The main contribution to the $ep$ cross section originates from
the region of virtualities $q_\bot^2/(1-y)+q_e^2$ much lower than
the characteristic scale of hadronic interactions. Therefore, {\it
(i)} the distribution is peaked at very small $q_\bot$ when the
scattered electron escapes observation, {\it (ii)} with high
enough accuracy one can take for the amplitudes of $\gamma p$\ \
subprocess their on-shell values discussed above, {\it (iii)} the
precision of eq.~(\ref{WWflux}) is very high (much better than
$k_{\bot,max}^2/m_\rho^2$). At this transition from photons to
electrons, the quantities (\ref{asyms}) that describe the charge
asymmetry are transformed as follows: the FB variable $\xi$ stays
unchanged since it is independent of small change of transverse
momentum and keeps its form under the longitudinal boost; the
transverse variable $v$ will be now defined by the same expression
(\ref{asyms}), but in terms of the transverse dipion momentum
$\vec K_\bot = \vec q_\bot + \vec k_\bot$ with respect to $ep$
collision axis.

The differential cross section of dipion production in $ep$
collisions is given by convolution of flux (\ref{WWflux}) with
(\ref{sigtot}), (\ref{dif1}). For numerical estimates  in our
kinematical region (\ref{def0}) it is useful to change the
$k_\bot^2$ dependence from (\ref{rhocrsec}) to $1/(1+Bk_\bot^2)$
which is good approximation at considered $Bk_\bot^2<0.1$). In the
region (\ref{dom0}) we have also $K_\bot^2\gg Q_m^2,\,q_e^2$ and
the charge symmetric part of cross section can be written as (see
e.g. \cite{1975})
 \bear{c}
\dfrac{d\sigma^{ep}_{Pom}}{ dM^2\, d K_\bot^2\, d\xi\, dv\, dy} =
\dfrac{\beta}{  \sqrt{1-\xi^2-v^2}}
|A_{1,0}(s)|^2 |D_1(M^2)|^2 \dfrac{3}{4\pi} (1 -\xi^2) \\[5mm]
\cdot \dfrac{\alpha}{\pi
y}\left[\left(1-y+\dfrac{y^2}{2}\right)\cdot \left(\log\dfrac{1}{
Bq_e^2}-1\right) -
(1-y)\right]\,,\\[5mm]
\dfrac{d\sigma^{ep}_{Prim}}{ dM^2\, d K_\bot^2\, d\xi\, dv\, dy} =
\dfrac{\beta}{\sqrt{1-\xi^2-v^2}} \sigma_2
|D_2(M^2)|^2\\[4mm]
\cdot \dfrac{15}{ 8\pi} \left[\fr{g_2^2}{2}(1 -\xi^2)^2 +3g_0^2
\left(\xi^2-\fr{1}{3}\right)^2
 +g_0g_2\sqrt {6}\left(\xi^2-\fr{1}{3}\right)
(2v^2+\xi^2-1)\right]\\[4mm]
\cdot\dfrac{\alpha}{\pi y
(K^2_\bot+Q_m^2+q_e^2)}\left[\left(1-y+\dfrac{y^2}{2}\right) \cdot
\left(\log\left(\dfrac{(K_\bot^2)^2}{Q_m^2q_e^2}\right)-2\right)
-(1-y)\right]\nonumber \,.\end{array}\label{epf}
 \ee
Note that in the Primakoff contribution we also keep term
$Q_m^2+q_e^2$ in the denominator, which is negligible in the
considered kinematical range but useful in the estimate of total
cross section.

The charge asymmetric contribution is written now via new value
$v$ as
 \bear{c}
\dfrac{d\sigma_{asym}^{ep}}{dM^2\, dK_\bot^2\, d\xi\, dv\, dy} =
\dfrac{v\cdot\beta}{ \sqrt{1-\xi^2-v^2}} \cdot \dfrac{3\sqrt{5}}{
4 \pi} |A_{1,0}(s)| \sqrt{\sigma_2}  \cdot \left[g_2(1-\xi^2) +
g_0\sqrt{\dfrac{2}{ 3}}(3\xi^2-1)\right]
\\[4mm] \cdot \dfrac{\alpha |K_\bot|}{ \pi y K_\bot^2}
\left[\left(1-y+\dfrac{y^2}{2}\right) \cdot \left(
\log\dfrac{K_\bot^2}{q_e^2} - \dfrac{1}{2}\right) -
\dfrac{1-y}{2}\right]\cdot{\cal I}_{\rho f}(M^2)\,.
\label{difep}\end{array}
 \ee

The total values of the signal and background integrated over
entire region (\ref{dom0}) similar to those written in
eq-s~(\ref{bkgd1}),~(\ref{asym1}) and with the same notations,
written with logarithmic accuracy (for estimates), are
 \bea
\fr{\sigma^{ep}_{bkgd}}{dy}= N_\gamma(y) \left[\sigma_\rho B
\,C_1\ln\dfrac{1}{Bq_e^2}(K_{max}^2-K_{min}^2) +
\sigma_2\,C_2\ln\dfrac{K_{max}^2}{K_{min}^2}\ln
\dfrac{K_{max}^2K_{min}^2}{Q_m^2q_e^2} \right]\,,\nn\\
\label{finest}\\
\fr{\Delta\sigma_{chas,T}^{ep}}{dy} =  N_\gamma(y)\dfrac
{9\sqrt{5}}{8} \sqrt{\sigma_\rho B \cdot \sigma_2} \cdot
\Delta{\cal I}_{\rho f}\cdot \left(K_{max}\ln\dfrac{
K_{max}^2}{q_e^2}-K_{min}\ln\dfrac{ K_{min}^2}{q_e^2}\right)\,.\nn
 \eea
Here $N_\gamma(y)=(\alpha/\pi y)(1-y+y^2/2)$.

\section{Estimates of the effect}

\subsection{Extracting charge asymmetry}

For the integrated luminosity ${\cal L}$, the statistical
significance of the result is given by the ratio of the number of
events under interest ${\cal L} \Delta\sigma_{chas,T}$ to the
dispersion of background events $\sqrt{{\cal L}\,\sigma_{bkgd}}$,
 \be
 SS=\dfrac{{\cal L} \Delta\sigma_{chas,T}}{\sqrt{{\cal
 L}\,\sigma_{bkgd}}}\,.\label{ssdef}
 \ee
In particular, we consider {\it local statistical significance} $
SS(M)$ defined by this very equation for fixed value of dipion
mass $M$, $SS(M)={\cal L} d\Delta\sigma_{chas,T}/dM^2/\sqrt{{\cal
L}\, d\sigma_{bkgd}/dM^2}\;$. The study of shape of this $SS(M)$
helps us in the choice of cuts in $M$ for data processing.

\begin{figure}[!htb]
   \centering
\includegraphics[width=0.8\textwidth]{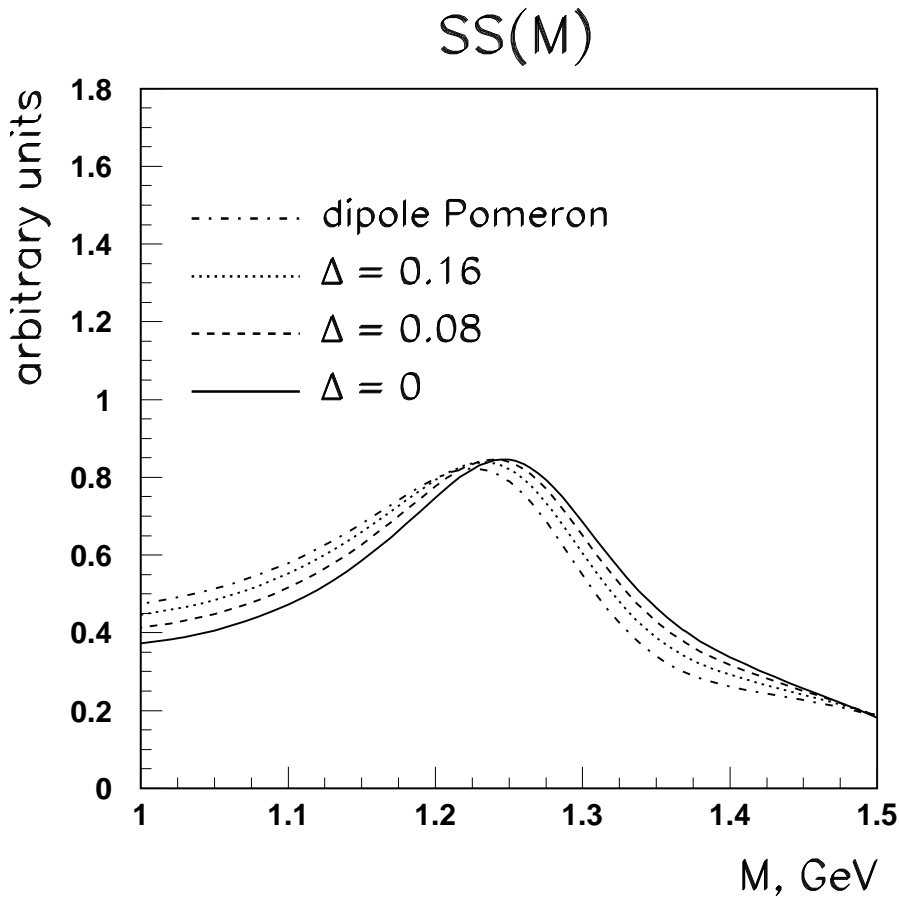}
\caption{\sf The local statistical significance of the charge
asymmetry (arbitrary units). The solid and dashed curves
correspond to $\Delta_F = 0$  and 0.16, respectively.}
   \label{figss}
\end{figure}

Fig.~\ref{figss} shows this local statistical significance.
For the $C$-odd dipions, as
said above, we assume the $\rho$-meson dominance with $|A_{1,0}|$
given by eq.~(\ref{rhocrsec}).  For $f_2$ meson production,
$\sigma_2$ in (\ref{agamma}) is $\sigma_2=0.42$ nb. Besides, we
set $g_0=0$, $g_2=1$ in accordance with data for $f_2$ production
in photon collisions. All other parameters were already discussed.

Let us remind that the diffractive dipion photo-production
dominates over Primakoff contribution in the C--even part of cross
section. Therefore, in the mass region, where $f_2$ production
dominates in Primakoff effect, the local statistical significance
is estimated as
 $$
 SS(M)\propto Re(D_2^*e^{i\delta_F}D_1)/|D_1| \leq |D_2(M)|
 $$
This suggests that the largest statistical significance comes
approximately from the region under the $f_2$ meson peak $m_f -
\Gamma_f < M < m_f + \Gamma_f$. This is the reason why we study
here the charge asymmetry only in the region $1.1 < M < 1.4$ GeV.

Integration of $|D_i|^2$ and of the overlap function ${\cal
I}_{\rho f}$ over this range gives for quantities in
\eqref{finest} at $\Delta_F=0$
 \bear{c}
 \Delta{\cal I}_{\rho f} = \int dM^2 {\cal I}(M^2) = 0.114\,;\\[3mm]
 C_1 = \int dM^2 |D_1(M^2)|^2 = 0.045\,;\quad
 C_2 = \int dM^2 |D_2(M^2)|^2 = 0.36\,.
 \end{array}\label{ci}
 \ee
Note that the value of $\Delta{\cal I}_{\rho f}$ depends on
$\Delta_F$ only weakly.

\subsection{ Numerical estimates}

$\bullet$ {\large\bf\bm $\gamma p$ collisions.} Now the
statistical significance of observation of charge asymmetry
(\ref{ssdef}) is evidently independent of the upper cut $k_{max}$.
The upper cut $k_{max} = 100$ MeV guarantees that the odderon
contribution is negligible. The resulting cross sections are
 \be
 \sigma_{bkgd} = 49 \mbox{ nb}\,;\quad \Delta\sigma_{chas,T} = 5.2 \mbox{ nb}\,;\quad
 SS_T \approx 24\cdot\sqrt{{\cal L}_{\gamma p}(\mbox{pb}^{-1})}\,.
 \ee

$\bullet$ {\large\bf  \boldmath $ep$ collisions.} For the $ep$
collisions, we take, for definiteness, ${\cal L}_{ep}=100$
pb$^{-1}$ and integrate over the $y$-interval $0.2 < y < 0.8$. We
then obtain the following values of the cross sections and of the
statistical significance
 \be
\sigma_{bkgd}^{ep} \approx 1.5\mbox{ nb}\,,\quad
\Delta\sigma_{chas,T}\approx 0.13\mbox{ nb}\, \quad \Rightarrow
\quad SS_T \approx 34\,.\label{SS}
 \ee

{\large\bf\boldmath Sensitivity to $\delta_F$}. The above values
of the integral $SS$ show that the effect is observable at HERA
with good confidence. We hope that after dedicated specification
of the models for $D_i$, a detailed study of the $M$-shape of this
charge asymmetry will allow for extraction of the Pomeron phase
$\delta_F$ with reasonable precision.

\section{Discussion and conclusions}

We showed that the interference between the Pomeron exchange and
the Primakoff effect contributions gives charge asymmetry in pion
distributions. The absolute value and the shape of $M$--dependence
of this charge asymmetry is sensitive to the phase of the strong
amplitude (the Pomeron phase) $\delta_F$. Accurate study of this
shape can lead to a direct measurement of $\delta_F$. Our
estimates show that this effect can be studied at HERA.

The approach suggested avoids the problems associated with the
measurement of very small transverse momenta of the detected
particles, in contrast to the strong--Coulomb interference in
elastic $pp$ scattering (where one should measure transverse
momenta\linebreak[4]  $p_\bot\lesssim 100$ MeV). Here, detected
pions have typical transverse momenta $|p_{\pm \bot}|\sim 500$ MeV
(for higher $M$), which looks measurable better than proton
momenta in the mentioned case of Coulomb interference.

Equations written in the text allow one to obtain preliminary
estimate for $\delta_F$ and find its $s$ dependence with accuracy
limited by details of experimentation. A more precise extraction
of the absolute value of $\delta_F$ demands more accurate models
both for Pomeron and Primakoff amplitudes. The main features of
these models are well known, and these models can be further
improved right in the course of dedicated experiments on charge
asymmetries (both at high-energy lepton-hadron or low-energy
$e^+e^-$ colliders). The sketch of how predictions can be made
more precise is given in the text. The invariant mass interval
$M = 1.1 \div 1.3$ GeV seems to be particularly suitable, since
theoretical predictions can be made more precise here. For each
mass interval, these problems should be studied separately.

\bu {\bf\bm The case $M < 1$ GeV}. At lower dipion invariant
masses, $M \lsim 1$ GeV, the study of the transverse charge
asymmetry can also be used for extraction the Pomeron phase. A
more detailed model for the $\gamma\gamma\to \pi^+\pi^-$ reaction
is necessary to make more accurate predictions for the study of
the Pomeron phase. This model can be verified by measurement of
similar charge asymmetry in the process $e^+e^-\to
e^+e^-\pi^+\pi^-$ at modern $e^+e^-$ colliders \cite{GSS2001}.
That is the subject of forthcoming studies.

Preliminary estimates show that below the $\rho$ peak the phases
of factors $D_1$ and $D_0$ are close to each other, so that the
contribution of their interference term to the considered symmetry
is small. The dominant contribution to the charge asymmetry is
given here by the $\rho/QED$ interference. The best statistical
significance of charge asymmetry comes from the region $M=0.4-0.8$
GeV.

\bu The extension of this idea to nuclear targets is
straightforward. A detailed treatment of charge asymmetry in
dipion production in
$e A$ collisions (e-RHIC or nuclear LHC) will be given elsewhere.\\

We are thankful to A.B. Kaidalov, N.N.~Nikolaev, D.Yu.~Ivanov,
V.V.~Serebryakov, V.G.~Serbo and G.N.~Shestakov for valuable
comments. IPI is also thankful to Prof.~R.~Fiore for hospitality
at INFN, Cosenza, where the paper was finished. This work was
supported by INTAS grants 00-00679 and 00-00366, RFBR grant,
NSh-2339.2003.2.


\begin{thebibliography}{99}



\bibitem{pp} E-811 Coll., {\em Phys. Lett.} {\bf B537}, 41 (2002).



\bibitem{jenk}
R.~Fiore, L.~Jenkovszky, F.~Paccanoni and A.~Papa, {\em Phys.
Rev.} {\bf D65}, 077505 (2002); L.~L.~Jenkovszky, E.~S.~Martynov and F.~Paccanoni,
talk given at HADRONS 96, Ukraine, 9-16 Jun 1996, hep-ph/9608384.



\bibitem{Prokudin}
V.~A.~Petrov, E.~Predazzi and A.~Prokudin,
{\em Eur. Phys. J.} {\bf C28}, 525 (2003).


\bibitem{chargereview} I.F.~Ginzburg,
``Charge asymmetry in $e^\pm e^-$, $ep$, $e\gamma$, $\gamma\gamma$
collisions'', {\em hep-ph/0211099}.



\bibitem{GIN2001}
I.F.~Ginzburg, I.P.~Ivanov and N.N.~Nikolaev, {\em Eur.\ Phys.\
J.\ direct} {\bf C5}, 02 (2003); talk given at DIS 2001, Bologna,
Italy, 27 Apr -- 1 May 2001, {\em hep-ph/0110181}.

\bibitem{GIv2004} I.F.~Ginzburg, I.P.~Ivanov, {\em Phys.
Elem.part. Atomic Nucl.} {\bf 35} (2004) 29-45.



\bibitem{DM2} D.~Bisello {\it et al.}  [DM2 Collaboration],
{\em Phys. Lett.} {\bf B220}, 321 (1989).



\bibitem{f2shape}
J.~Boyer {\it et al.}, {\em Phys. Rev.} {\bf D42}, 1350 (1990).



\bibitem{gamgampipi} M.~Boglione and M.R.~Pennington,
{\em Eur.\ Phys.\ J.} {\bf C9}, 11 (1999).



\bibitem{1975}V.M.~Budnev, I.F.~Ginzburg, G.V.~Meledin and V.G.~Serbo,
{\em Phys.\ Rept.} {\bf 15}, 181 (1975).



\bibitem{GSS2001} I.F.~Ginzburg, A.~Schiller and V.G.~Serbo,
{\em Eur.\ Phys.\ J.} {\bf C18}, 731 (2001).



\end{thebibliography}
\end{document}